\documentclass[rnote]{aa} 
\usepackage{graphicx}
\usepackage{txfonts}
\newcommand{\recta}[3]{$#1^{\rm{h}}\,#2^{\rm{m}}\,#3^{\rm{s}}$}
\newcommand{\dec}[3]{$#1\degr\,#2\arcmin\,#3\arcsec$}

\begin{document}
\newcommand{\kms}{km\,s$^{-1}$}

  \title{CO in OH/IR stars close to the Galactic centre}

  \titlerunning{CO in OH/IR stars at the Galactic centre}

  \author{A.~Winnberg  \inst{1}
     \and S.~Deguchi   \inst{2}
     \and M.J.~Reid    \inst{3}
     \and J.~Nakashima \inst{4,} \inst{5}
     \and H.~Olofsson  \inst{1,} \inst{6}
     \and H.J.~Habing  \inst{7}}

  \offprints{A.~Winnberg}

  \institute{Onsala Space Observatory, Observatoriev\"{a}gen,
             SE--439 92 Onsala, Sweden\\
             \email{anders.winnberg@chalmers.se}
        \and Nobeyama Radio Observatory, Minamisaku, 
             Nagano 384--1305, Japan
        \and Harvard-Smithsonian Center for Astrophysics, 60 Garden Street,
             Cambridge MA 02138, USA
        \and Academia Sinica Institute of Astronomy and Astrophysics, P.O. Box
             23--141,Taipei 106, Taiwan
        \and Department of Physics, University of Hong Kong, 
             Pokfulam Road, Hong Kong
        \and Stockholm Observatory, AlbaNova University Centre,
             SE--106 91 Stockholm, Sweden
        \and Sterrewacht Leiden, P.O. Box 9513,
             NL--2300 RA Leiden, The Netherlands}

  \date{Received date; accepted date}

 \abstract
  {}
{A pilot project has been carried out to measure circumstellar CO emission from
three OH/IR stars close to the Galactic centre.  The intention was to find out
whether it would be possible to conduct a large-scale survey for mass-loss rates
using, for example, the Atacama Large Millimeter Array (ALMA).  Such a survey
would increase our understanding of the evolution of the Galactic bulge.}
{Two millimetre-wave instruments were used: the Nobeyama Millimeter Array at
115~GHz and the Submillimeter Array at 230~GHz.  An interferometer is necessary
as a `spatial filter' in this region of space because of the confusion with
\textit{inter}stellar CO emission.}
{Towards two of the stars, CO emission was detected with positions and
radial velocities coinciding within the statistical errors with the
corresponding data of the associated OH sources.  However, for one of the stars
the line profile is not what one expects for an unresolved expanding
circumstellar envelope.  We believe that this CO envelope is partially resolved
and that this star therefore is a foreground star not belonging to the bulge.}
{The results of the observations have shown that it is possible to detect line
profiles of circumstellar CO from late-type stars both within and in the
direction of the Galactic bulge.  ALMA will be able to detect CO emission in
short integrations with sensitivity sufficient to estimate mass-loss rates from
a large number of such stars.}


  \maketitle
%

\section{Introduction}

Rapid stellar mass loss occurs at the so-called red giant and, in
particular, the asymptotic giant branches (RGB and AGB).  Due to high stellar
density and rapid star formation, the Galactic bulge contains large numbers of
RGB and AGB stars.  It is important to measure the mass-loss rates of these
stars for a picture of the recirculation of matter and metal enrichment in this
region of the Milky Way.  In addition, stars in the central bulge are at
distances that differ from one another by merely a few percent leading to an
accurate estimate of the mass-loss-rate
\textit{distribution}.

The most accurate method of estimating the stellar mass-loss rates is based on
the CO rotational spectral lines (cf. Ramstedt et al., \cite{Ramstedt08}).  The
first attempt to detect such circumstellar lines in the vicinity of the Galactic
centre (GC) was made by Mauersberger et al. (\cite{Mauersberger88}) using the
IRAM 30-m telescope.  They detected the $J$=1$\rightarrow$0 and 2$\rightarrow$1
CO lines from the proto-planetary nebula (PPN) OH0.9+1.3 which has a high radial
velocity \linebreak(--110~\kms) making it likely to be physically close to the
GC.  They also proposed to look for CO emission from OH/IR stars, since such
stars are believed to be the progenitors of PPNe.  Winnberg et al.
(\cite{Winnberg91}) used the same radio telescope to observe several OH/IR stars
close to the GC but detected CO emission from only one star: OH0.3--0.2 (Baud et
al., \cite{Baud75}). This star has a very high radial velocity (--341~\kms) and
therefore the circumstellar emission is not affected by the interstellar
background emission that covers typically the velocity range --200 to +200~\kms.
 None of the other candidate stars were detected because of confusion with
interstellar emission.

The present project employs a different observing technique in an attempt to
detect the \textit{circum}stellar CO emission in the midst of
\textit{inter}stellar CO emission.  A radio interferometer with suitable
baselines can be used as a `spatial filter' by resolving most of the
interstellar background but leaving the circumstellar emission as unresolved
point sources.

To determine optimal baseline lengths and the most favourable CO lines,
we started a pilot experiment.  We chose three OH/IR stars close to the position
of the GC with strong IR fluxes and with low-to-moderate radial velocities.  In
2003 -- 2004 we used the Nobeyama Millimeter Array (NMA) at 115~GHz (CO,
$J$=1$\rightarrow$0) and in 2005 we used the SubMillimeter Array (SMA; Mauna
Kea, Hawaii) at 230~GHz ($J$=2$\rightarrow$1).

This research note presents the main results of these two data sets and outlines
the prospects for future systematic surveys of late-type stars in the Galactic
bulge using the Atacama Large Millimeter Array (ALMA).  A conference report of
this project appears in Winnberg et al. (\cite{Winnberg06}).


\begin{figure}[ht]
\centering
\includegraphics[width=18pc]{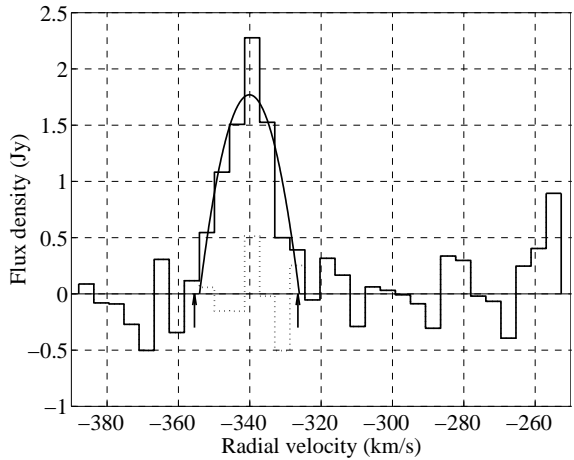}
\caption{\label{fig:TestStarSpec} Circumstellar CO(2$\rightarrow$1) line profile
of OH0.3--0.2.  The thick `staircase' line is the observed spectrum; the thick
curved line is a fitted parabola; the dotted line depicts the residuals (i.e.,
observed spectrum minus parabola).  The arrows indicate the velocities of OH
masers.}
\end{figure}

\begin{figure}[ht]
\centering
\includegraphics[width=18pc]{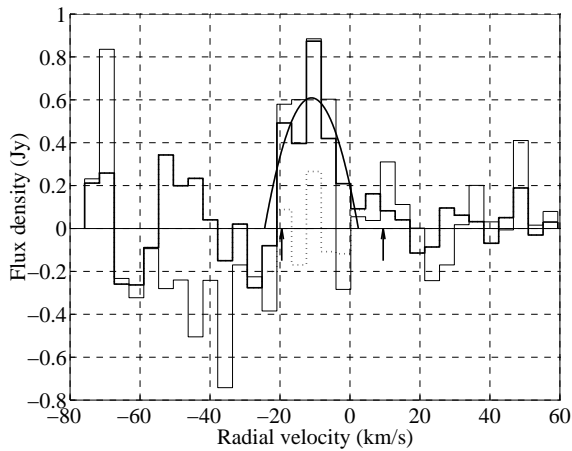}
\caption{\label{fig:OH359.762+0.120Spec} Circumstellar CO(2$\rightarrow$1) line
profiles of OH359.762+0.120.  The thick `staircase' line is the observed
spectrum with baseline lengths shorter than 25~k$\lambda$ excluded.  The
parabola, the residuals and the OH maser velocities are represented as in Fig.
\ref{fig:TestStarSpec}. In addition, a thin `staircase' line indicates an
observed spectrum with all projected baseline lengths included.}
\end{figure}

\section{Observations}
The first observations in this project were done with the 6-element array (NMA)
at Nobeyama Radio Observatory, Japan, in November 2003 and in January 2004 at
115~GHz.  Three OH/IR stars were selected for observations on the basis of
strong IR fluxes: OH359.117--0.169, OH359.762+0.120 and OH359.971--0.119 (Ortiz
et al., \cite{Ortiz02}).  The relevant data for these stars are listed in Table
\ref{tab:OH/IRstars}, where $F_{[15]}$ denotes the flux density at a wavelength
of 15~$\mu$m as measured by the Infrared Space Observatory (ISO).  A clear
signal was obtained from OH359.971--0.119 (Fig. \ref{fig:OH359.971-0.119Spec},
dotted line).  The position  of the CO line agreed with the position of the
1612-MHz OH line.  However, the line width was too narrow for an unresolved
circumstellar CO source and the radial velocity of the line was close to that of
the red-shifted OH line component rather than to the average velocity of the two
OH components.

Data were taken using two digital correlation spectrometers simultaneously.  One
of them consisted of 128 channels with a width of 8.0~MHz.  The other one had
1024 channels of width 31.25~kHz.

\begin{table*}[ht]
\begin{minipage}[t]{18cm}
\caption[]{Selected OH/IR Stars}
\label{tab:OH/IRstars}
\begin{center}
\begin{tabular}{lccrrrc}
\hline\noalign{\smallskip}
Name             &    RA                 &   Dec               &   $V_{\rm rad}$
& $V_{\rm exp}$ & $F_{[15]}$          & Ref. \\
                & (J2000)               & (J2000)             & (\kms)         
& (\kms)        & (Jy)         &      \\
\noalign{\smallskip} \hline\noalign{\smallskip}
OH0.3--0.2       & \recta{17}{47}{06.95} & \dec{-28}{44}{42.2} & --341.0 & 14.5 
&      & 1  \\
OH359.117--0.169 & \recta{17}{47}{21.79} & \dec{-29}{47}{42.2} &  --88.5 & 21.2 
& 6.9 & 2  \\
OH359.762+0.120  & \recta{17}{44}{34.95} & \dec{-29}{04}{35.2} &   --5.7 & 15.3 
& 13.8 & 3  \\
OH359.971--0.119 & \recta{17}{46}{00.94} & \dec{-29}{01}{23.6} &   --8.5 & 19.3 
& 9.5 & 3  \\
\noalign{\smallskip} \hline
\end{tabular}
\begin{flushleft}
{\bf References:}
1.--Fix \& Mutel (\cite{Fix84}); 2.--Sevenster et al. (\cite{Sevenster97});
3.--Lindqvist et al. (\cite{Lindqvist92})
\end{flushleft}
\end{center}
\end{minipage}
\end{table*}

\begin{table*}[ht]
\begin{minipage}[t]{18cm}
\caption[]{Results}
\label{tab:results}
\begin{center}
\begin{tabular}{lccccccc}
\hline\noalign{\smallskip}
Name             & Trans.          &    RA                                      
&   Dec                    &  $V_{\rm m}$ & $V_{\rm e}$ & $S_{\rm m}$ &
$\dot{M}$   \\
                & ($J$)           & (J2000)                                    
& (J2000)                  & (\kms)       & (\kms)      & (Jy) &
($\rm {M}_\odot$\,yr$^{-1}$)  \\
\noalign{\smallskip} \hline\noalign{\smallskip}
OH0.3--0.2       & 2$\rightarrow$1 & \recta{17}{47}{06.978}(0.009)              
& \dec{-28}{44}{42.8}(0.2) & --340.1(0.6) & 14.0(0.8)   & 1.8(0.1) & $5 \times
10^{-4}$ \\
OH359.117--0.169 & 1$\rightarrow$0 &                                            
&                          &              &             &$\lesssim$0.2 &  \\
                & 2$\rightarrow$1 &                                            
&                          &              &             &$\lesssim$0.54 & \\
OH359.762+0.120  & 1$\rightarrow$0 &                                            
&                          &              &             &$\lesssim$0.15 & \\
                & 2$\rightarrow$1 & \recta{17}{44}{34.979}(0.012)              
& \dec{-29}{04}{36.6}(0.2) &  --11(1)     & 13(1)       & 0.6(0.1) & $4 \times
10^{-4}$ \\
OH359.971--0.119 & 1$\rightarrow$0 &
\recta{17}{46}{00.75}\hspace{1.8mm}(0.04)\hspace{1.3mm} &
\dec{-29}{01}{21.5}(0.7) &              &             & $\sim$0.4  &   \\
                & 2$\rightarrow$1 & \recta{17}{46}{00.909}(0.008)              
& \dec{-29}{01}{23.2}(0.2) &              &             & $\sim$5  &     \\

\noalign{\smallskip} \hline
\end{tabular}
\end{center}
\end{minipage}
\end{table*}

In June 2005 the same three stars plus the previously detected star OH0.3--0.2
were observed using the 8-element array (SMA) on Mauna Kea, Hawaii, at 230~GHz. 
The spectrometer was configured to give a resolution of 3.25~MHz.  The test star
OH0.3--0.2 was detected and showed properties in accordance with the single-dish
data (Fig. \ref{fig:TestStarSpec}).  In addition the stars OH359.762+0.120 and
OH359.971--0.119 were detected (Figs \ref{fig:OH359.762+0.120Spec} and
\ref{fig:OH359.971-0.119Spec}).

Typical synthesized beam FWHM values of images produced with all baselines were
4\arcsec\ $\times$ 7\arcsec\ for the NMA and 3\arcsec\ $\times$ 4\arcsec\ for
the SMA.


\section{Data reduction}
\subsection{The NMA}
The data were calibrated using the Nobeyama internal programme package and
subsequently the Astronomical Image Processing System (AIPS) was used in a
standard way to obtain images and spectra.

We found strong ripples in the images that could be eliminated by removing data
from projected baselines shorter than 10~k$\lambda$ (26~m).  Due to atmospheric
phase instability, we decided to remove all projected baselines longer than
40~k$\lambda$ (104~m) and to introduce a gaussian baseline-length taper such
that baselines of length 40~k$\lambda$ got a weight of 30~\%.

Maps were made with $256 \times 256$ pixels of size 0.5$\arcsec$ with uniform
weighting.   They were 'cleaned' using the standard H\"{o}gbom/Clark algorithm
with a gain of 0.1 and a minimum flux density per clean component being the
product of the beam dynamic range ($1/|\rm{strongest\,sidelobe}|$) and the
expected \textit{rms} noise fluctuations.  This ensured that no `overcleaning'
took place in the rather noisy maps.

Data cubes were made consisting of images from the 20 central channels of the
low-resolution spectrometer covering 416~\kms.  For the high-resolution
spectrometer similar data cubes were made consisting of 256 channels where each
channel resulted from averaging 4 original channels.  Therefore these spectra
covered 83.2~\kms with a resolution of 0.325~\kms.

\subsection{The SMA}
Data from this instrument were retrieved using software from the Radio Telescope
Data Center (RTDC) of the Center for Astrophysics (CfA).  Calibration was
performed using the image processing package IDL-MIR at Academia Sinica
Institute of Astronomy and Astrophysics (ASIAA).  Further data reduction was
made using both AIPS at Onsala Space Observatory (OSO) and Miriad (SMA version)
at ASIAA.

The {\it uv} data were investigated for the presence of spatially extended
emission by plotting the visibility amplitude as a function of projected
baseline length.  Based on such plots it was decided to exclude baselines
shorter than 25~k$\lambda$ (32.5~m) for OH359.762+0.120 and shorter than
30~k$\lambda$ (39~m) for OH359.971--0.119 in order to avoid, as far as
reasonable, contamination by residuals of interstellar emission.  For
OH359.117--0.169 no evidence of significant interstellar emission was found. 
No interstellar emission was found in the IF band of the test source OH0.3--0.2
as expected.

Map-cubes were made with 256 $\times$ 256 pixels of size 0.5\arcsec, however,
this time with natural weighting.  They were 'cleaned' using the standard
H\"ogbom/Clark algorithm in a manner similar to the treatment of the NMA data. 
When a compact source was seen in one of the channels within the OH velocity
span at a position close to the position of the OH source, we fitted a
two-dimensional elliptical Gaussian and extracted a spectrum at the pixel
closest to this position.

After exclusion of the shorter baseline lengths all the maps are free from
interstellar CO emission except for occasional point sources.  We interpret them
as unresolved rests of interstellar emission.  Since most of the emission is
resolved there are `negative point sources' as well, i.e. unresolved `dips' in
the interstellar background.  The positions of the OH/IR stars are known to
1\arcsec\ or better, and therefore there is never any doubt about the
identification of a circumstellar CO source.  In addition to the position
evidence, there is the radial velocity evidence that strengthens the
identification case quite considerably.


\begin{figure}[ht]
\centering
\includegraphics[width=18pc]{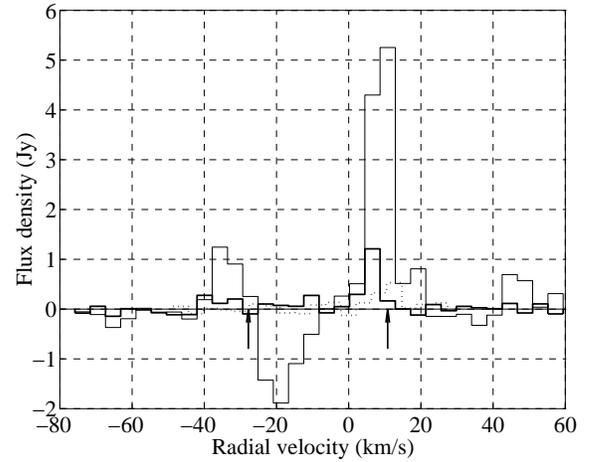}
\caption{\label{fig:OH359.971-0.119Spec}Circumstellar CO(2$\rightarrow$1) line
profiles of OH359.971--0.119.  The definitions of the lines are given in the
caption of Fig. \ref{fig:OH359.762+0.120Spec}, except in this figure the dotted
`staircase' line is the CO(1$\rightarrow$0) spectrum as observed with the NMA.}
\end{figure}

\section{Results}
Table \ref{tab:results} lists the measured parameters of the CO sources
associated with the four stars observed.  The 1-$\sigma$ errors are given in
parentheses after the values.  For OH0.3--0.2 and OH359.762+0.120
(2$\rightarrow$1) least-square fits of parabolas to the CO line profiles have
been made, assuming that the data are from unresolved, optically thick emission:

\begin{equation}
S = S_{\rm m}\left[ 1 - \left( \frac{V - V_{\rm m}}{V_{\rm e}}\right)^2\right] 
\end{equation}

\noindent where $S_{\rm m}$ is the maximum flux density at the radial velocity
$V_{\rm m}$, and $V_{\rm e}$ is half the line width at zero intensity.  These
three parameters are given in the table together with the statistical errors
from the fitting procedure.  No CO emission associated with OH359.117--0.169
(1$\rightarrow$0 and 2$\rightarrow$1) and OH359.762+0.120 (1$\rightarrow$0) was
found and 3-$\sigma$ upper limits are given for $S_{\rm m}$.  The CO line
profiles for OH359.971--0.119 (1$\rightarrow$0 and 2$\rightarrow$1) are
incompatible with a parabola, although the CO positions are coincident with the
OH position, and only approximate values of the maximum flux densities are given
(Fig. \ref{fig:OH359.971-0.119Spec}).

OH359.117--0.169 was not detected at any of the two CO lines and we do not know
the reason for it.  One guess would be that this is due to heavy absorption by
interstellar CO clouds in front of the star.  Such a case could possibly occur
for a star that is situated at a distance beyond the GC.

OH359.762+0.120 was detected with a rather poor S/N.  Therefore there are quite
large errors associated with the elements of the fitted parabola (Fig.
\ref{fig:OH359.762+0.120Spec} and Table \ref{tab:results}).  Within these
errors, the mean radial velocity of the line and the line width are compatible
with the systemic velocity of the star and its envelope expansion velocity as
measured from the OH line profile.  Notice that the exclusion of short baselines
did not improve the detection of the line significantly.  It merely improved the
spectral baseline.

OH359.971--0.119 was detected with moderate to good S/N.  However, the CO line
is narrow and close to the 'red-shifted' OH line component, i.e. the backside of
the envelope (Fig. \ref{fig:OH359.971-0.119Spec}).  A similar, but considerably
weaker, line was observed with the NMA at 115~GHz (dotted line).  Notice that
the line profile obtained when all the {\it uv} data were included (thin solid
line) shows a much stronger and broader `red-shifted' line and even a
`blue-shifted' counterpart.  The `negative signal' near $-20$~\kms\ is probably
caused by a `dip' in the interstellar background, as discussed above, but in
this case it is resolved.

Mass-loss rates have been calculated for OH0.3--0.2 and OH359.762+0.120
(2$\rightarrow$1) using a new formula based on the original equation by Knapp \&
Morris (\cite{Knapp85}) but containing constants determined by least-squares
fits to physical models (Ramstedt et al., \cite{Ramstedt08}), and the values
have been entered in Table \ref{tab:results}.  For these calculations a distance
of 8~kpc to the GC (Reid, \cite{Reid93}) and a $\rm CO/H_2$ abundance ratio of 
$2 \times 10^{-4}$ (Ramstedt et al., \cite{Ramstedt08}) have been assumed for
both stars.  The values of $V_{\rm e}$ have been taken as the expansion
velocities of the CSEs.  Both mass-loss rates are normal for OH/IR stars (we
caution that the validity of the Ramstedt et al. formula has only been tested in
the mass-loss-rate range 10$^{-7}$ to 10$^{-5}\,\rm {M}_\odot$\,yr$^{-1}$).


\section{Discussion}
\subsection{OH359.762+0.120}
There is little doubt, in spite of the poor S/N, that the detected CO source is
a true circumstellar source associated with the OH/IR star.  Because of the high
bolometric magnitude of this star and its strong OH emission, some people doubt
the association of this star with the Galactic bulge (see discussion by
Blommaert et al., \cite{Blommeart98}).  However, as pointed out by Blommaert et
al. (\cite{Blommeart98}), the OH radiation is heavily scattered by interstellar
free electrons (Frail et al., \cite{Frail94}) making it very likely that it
resides close to the GC.  Our result of a weak CO source leading to a normal
mass-loss rate for an assumed distance of 8~kpc supports this conclusion.

\subsection{OH359.971--0.119}
This star (as well as OH359.762+0.120) has detected 43-GHz SiO emission
(Lindqvist et al., \cite{Lindqvist91}).  However, the $v$=2, $J$=1$\rightarrow$0
SiO line is much stronger than its $v$=1 counterpart.  This suggests a cool dust
temperature and a high mass-loss rate (Nakashima \& Deguchi,
\cite{Nakashima07}).

The CO source associated with this star, on the other hand, is an enigma.  The
line profile is similar to that expected from a well-resolved circumstellar
envelope where unresolved emission is left at the front and back sides (see for
example the central CO source in U~Cam, Lindqvist et al., \cite{Lindqvist99}). 
For such a picture to be true, the star needs to be at a small distance and
would not belong to the Galactic bulge.  For example, assuming that the diameter
of the CO envelope is $2\,\times\,10^{17}$~cm (which might be an overestimate
for the 2$\rightarrow$1 transition) and that the angular diameter is 10\arcsec,
the distance would be only about 1.4~kpc.  The IR properties of this star also
are such that it is arguable whether it belongs to the bulge (cf. Ortiz et al.,
\cite{Ortiz02}).

Another explanation of the line profile would be that it is heavily distorted
through strong absorption of interstellar CO in front of the source.  A third
possibility - although improbable - is that the physical conditions in this CSE
are such that they favour weak maser action along radial directions (Morris,
\cite{Morris80}). Finally, there remains the (improbable) explanation that the
source is an unresolved remnant of interstellar CO emission that happens to be
at the same position and radial velocity as OH359.971--0.119.

Observations of higher-energy CO lines might be a possible way of revealing the
true nature of this source.  However, given the available data, we favour the
first alternative, i.e. that we have found a relatively nearby OH/IR star whose
CO envelope is resolved by both arrays used.  The salient points that support
this conclusion are:
\begin{itemize}
\item There are two CO line components whose positions and radial velocities
correspond to those of the OH line components.
\item The redshifted CO line component grows stronger and the blueshifted line
component emerges when all baselines (including the short ones) are included in
the imaging.
\item There is a weak line component at 115~GHz whose radial velocity coincides
with that of the redshifted 230-GHz component and whose position coincides with
that of the OH/IR star.
\item The 115-GHz line is much weaker than the 230-GHz counterpart which mostly
is the case for circumstellar CO lines in OH/IR stars especially those with cool
CSEs (Heske et al., \cite{Heske90}).
\end{itemize}

\section{Conclusions}
Our pilot project has shown that it is possible to detect circumstellar CO
envelopes of OH/IR stars close to the GC: out of three stars selected, one was
detected at both 115 and 230~GHz (OH359.971--0.119) and another one was detected
at 230~GHz only (OH359.762+0.120).  OH359.971--0.119 probably does not belong to
the Galactic bulge, but this fact is irrelevant for the issue at stake -- this
star too would have been hard to detect using a single-dish telescope.

We would have liked to observe the same three stars at 345~GHz
($J$=3$\rightarrow$2) and to try out somewhat longer baselines, but both these
requests require excellent atmospheric conditions and therefore the competition
for observing time is strong.

We have no doubt that ALMA will be able to detect a large number of OH/IR stars
in the inner bulge (Olofsson, \cite{Olofsson08}).


\begin{acknowledgements}
AW thanks the National Radio Observatory of Japan for a Visiting Professorship
during the observations at Nobeyama.  The NMA is operated by Nobeyama Radio
Observatory, a branch of National Astronomical Observatory of Japan.  The SMA is
a joint project between the Smithsonian Astrophysical Observatory and the
Academia Sinica Institute of Astronomy and Astrophysics and is funded by the
Smithsonian Institution and the Academia Sinica.
\end{acknowledgements}


\begin{thebibliography}{}

\bibitem[1975]{Baud75} Baud, B., Habing, H.J., O'Sullivan, J.D., Winnberg, A.,
Matthews, H.E., 1975, Nature, 258, 406
\bibitem[1998]{Blommeart98} Blommaert, J.A.D.L., van der Veen, W.E.C.J., van
Langevelde, H.J., Habing, H.J., Sjouwerman, L.O., 1998, A\&A, 329, 991
\bibitem[1984]{Fix84} Fix, J.D., Mutel, R.L., 1984, Astron. J., 89, 406
\bibitem[1994]{Frail94} Frail, D.A., Diamond, P.J., Cordes, J.M., van
Langevelde, H.J., 1994, ApJ, 427, L43
\bibitem[1990]{Heske90} Heske, A., Forveille, T., Omont, A., van der Veen,
W.E.C.J., Habing, H.J., 1990, A\&A, 239, 173
\bibitem[1985]{Knapp85} Knapp, G.R., Morris, M., 1985, ApJ, 292, 640
\bibitem[1999]{Lindqvist99} Lindqvist, M., Olofsson, H., Lucas, R., Sch\"{o}ier,
F.L., Neri, R., Bujarrabal, V., Kahane, C., 1999, A\&A, 351, L1
\bibitem[1991]{Lindqvist91} Lindqvist, M., Ukita, N., Winnberg, A., Johansson,
L.E.B., 1991, A\&A, 250, 431
\bibitem[1992]{Lindqvist92} Lindqvist, M., Winnberg, A., Habing, H.J., Matthews,
H.E., 1992, A\&AS, 92, 43
\bibitem[1988]{Mauersberger88} Mauersberger, R., Henkel, C., Wilson, T.L.,
Olano, C.A., 1988, A\&A, 206, L34
\bibitem[1980]{Morris80} Morris, M., 1980, ApJ, 236, 823
\bibitem[2007]{Nakashima07} Nakashima, J., Deguchi, S., 2007, ApJ, 669, 446
Stars'', eds. H.J. Habing \& H. Olofsson, Astronomy and Astrophysics Library,
Springer, New York, p. 325
\bibitem[2008]{Olofsson08} Olofsson, H., 2008, Astrophys. \& Space Sc., 313, 201
\bibitem[2002]{Ortiz02} Ortiz, R., Blommaert, J.A.D.L., Copet, E., Ganesh, S.,
Habing, H.J., Messineo, M., Omont, A., Schultheis, M., Schuller, F., 2002, A\&A,
388, 279
\bibitem[2008]{Ramstedt08} Ramstedt, S., Sch\"{o}ier, F.L., Olofsson, H.,
Lundgren, A.A., 2008, A\&A, 487, 645 (arXiv 0806.0517)
\bibitem[1993]{Reid93} Reid, M. J. 1993, ARA\&A, 31, 345
\bibitem[1997]{Sevenster97} Sevenster, M.N., Chapman, J.M., Habing, H.J.,
Killeen, N.E.B., Lindqvist, M., 1997, A\&AS, 122, 79
\bibitem[1991]{Winnberg91} Winnberg, A., Lindqvist, M., Olofsson, H., Henkel,
C., 1991, A\&A, 245, 195
\bibitem[2006]{Winnberg06} Winnberg, A., Deguchi S., Habing, H.J., Nakashima,
J., Olofsson, H., Reid, M.J., 2006, J. Phys. Conf. Ser., 54, 166

\end{thebibliography}
\end{document}